# Summary of The Superconducting RF Linac for Muon Collider and Neutrino Factory

J. Galambos (Oak Ridge National Laboratory), R. Garoby (CERN ), S. Geer (Fermi National Accelerator Laboratory)

Project-X is a proposed project to be built at Fermi National Accelerator Laboratory with several potential missions [1].  A primary part of the Project-X accelerator chain is a Superconducting linac, and In October 2009 a workshop was held to concentrate on the linac parameters [2]. The charge of the workshop was to "..focus only on the SRF linac approaches and how it can be used…".  The focus of Working Group 2 of this workshop was to evaluate how the different linac options being considered impact the potential realization of Muon Collider (MC) and Neutrino Factory  (NF) applications. In particular the working group charge was, " to investigate the use of a multi-megawatt proton linac to target, phase rotate and collect muons to support a muon collider and neutrino factory".  To focus the working group discussion, three primary questions were identified early on, to serve as a reference:

1) What are the proton source requirements for muon colliders and neutrino factories?
2) What are the issues with respect to realizing the required muon collider and neutrino factory proton sources?
   a. General considerations
   b. Considerations specific to the two linac configurations identified by Project-X.
3) What things need to be done before we can be reasonably confident that ICD1/ICD2 can be upgraded to provide the neutrino factory / muon collider needs?

A number of presentations were given, and are available at the workshop web-site [2]. This paper does not summarize the individual presentations, but rather addresses overall findings as related to the three guiding questions listed above.

## Baseline Linac Configurations

Project-X has already been under study for some years and studies have led to two primary accelerator configurations: "ICD-1" and "ICD-2" [3]. The Initial

Configuration Document (ICD)-1 design includes a pulsed (5 Hz) 8 GeV, 20 mA, 1.25 msec pulse length linac. The ICD-2 incorporates a CW, 1 mA, 2 GeV linac. These two linac configurations serve as a primary basis of comparison for the initial acceleration component for muon collider and neutrino factory applications for Project-X.

## Neutrino Factory / Muon Collider Requirements

The NF and MC proton driver beam requirements are summarized in Table 1 (taken from Ref. 4). The power requirement is based on MC luminosity specifications and neutrino delivery rates at far detectors. The required beam energy has a broad optimum centered around 10 GeV, with a steep fall-off below ~5 GeV (see Fig. 1 below, taken from Ref. 5). The bunch length of 2 nsec is a primary driver. Below a 1 nsec bunch length there is no benefit, but at 3 nsec there is a 10% loss in the muon production intensity. The NF and MC requirements differ in terms of number of bunches per pulse (1 for the MC versus 3-5 for the NF) and maximum repetition rate (15 Hz for the MC and 50 Hz for the NF). Producing a 4 MW proton beam in a single bunch at 15 Hz requires about 10 times the bunch intensity than in 3 bunches at 50Hz. Therefore we adopt the more stringent MC proton beam requirement as the basis of comparison for the mission here, as this will satisfy either mission.

Table 1. Proton driver requirements for a Neutrino Factory.

| Parameter | Value |
| --- | --- |
| Average beam power (MW) | 4 |
| Pulse repetition frequency (Hz) | $50^{a)}$ |
| Proton energy (GeV) | $10 \pm 5$ |
| Proton rms bunch length (ns) | $2 \pm 1$ |
| No. of proton bunches | 3 or 5 [b)] |
| Sequential extraction delay ($\mu$s) | $\geq 17$ |
| Pulse duration, liquid-Hg target ($\mu$s) | $\leq 40$ |
| Pulse duration, solid target (ms) | $\geq 20$ |

[a)] For a Muon Collider a lower repetition rate, 10–15 Hz, is required.
[b)] For a Muon Collider a single bunch is necessary

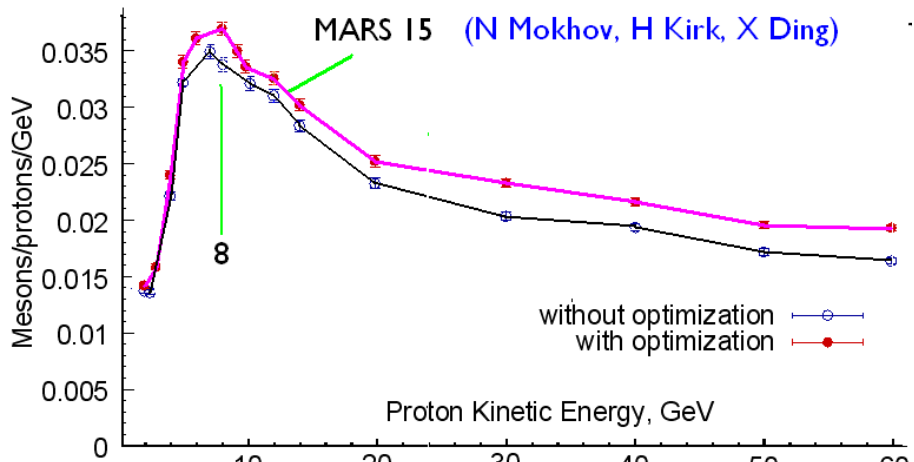
Figure 1 Meson production efficiency vs. proton beam energy (from Ref. 5).

## Proton Driver Issues

### General Issues

Achieving the required MC proton driver parameters is beyond existing accelerator capability and quite difficult, especially with respect to the short bunch structure. Some of the more challenging issues include:
- Keeping uncontrolled beam loss below 1 W/m (< 3 x $10^{-7}$) for hands on maintenance.
- Concentrating the collection of beam halo in a well-controlled collimation area (careful beam dynamics and detailed engineering is needed).
- Injection charge exchange stripping, issues such as foil heating, foil lifetime and beam loss from scattering during the multiple foil passages during injection.
- Maintaining a small longitudinal emittance, as required to achieve the short bunch length.
- Collective effects, in particular the large longitudinal phase space density makes longitudinal stability problematic.
- Large longitudinal compression factors required to reach the needed bunch length which necessitate RF gymnastics just before extraction.

## Potential Solutions

There are no fully developed NF/MC schemes available, although some promising linac based ideas were presented. In particular, there is not a complete set of beam and machine parameters for the different linac and rings associated with acceleration, accumulation and compression. We note that there is an historical RCS proton driver scheme at FNAL [6], but it does not meet the requirements specified above.

V. Lebedev [7] showed the difficulties in attaining the required MC parameters due to longitudinal microwave instability and tune shift due to transverse space charge. Using the ICD-2 driver alone the beam power is limited to < 1 MW. At 8 GeV, with some intermediate compression 1 MW is possible with a single bunch, and higher powers with multiple bunches compressed (as described below). Going to higher energy tends to alleviate the instability and space charge limitations.

Innovative ideas were introduced which tend to alleviate the compression and charge exchange injection issues. The first is the so-called "trombone/funnel" concept [5, 8] in which multiple 2 nsec bunches are created in a compression ring (each with a fraction of the intensity required by a single bunch 4 MW scheme), the individual bunches are extracted to separate extraction beam-lines of varying length so that they arrive at the target at the same time. Another new concept is a "resonant foil by-pass scheme" [8, 9], which reduces the foil traversal problem in the initial accumulation period for schemes that have very long injection times because of a low linac beam current. This idea involves chopping the CW linac beam to ~ 10% duty factor and incorporates a resonant position bump at the foil in the accumulation ring, synchronized so that the circulating beam passes through the foil during the un-chopped part of injection and is pulled away from the foil during the chopped portion of the injection. However details of this proposed scheme with realistic linac beam distributions and foil heating calculations have not been done.

One complete scheme was presented for a NF application [10], but this feasibility analysis was performed in the context of the CERN accelerator complex, and does not meet the MC requirements.

## General Comments

In order to highlight the specific areas needing attention for MC feasibility, the existing proposals should be further pursued. In general the linac based solutions (as opposed to RCS solutions) appear more promising. The relatively low energy (~ 10 GeV) for MC implementations results in long storage times to accumulate the required high intensity beams for the RCS based solutions (at even lower energies). The CW linac implementation (~ 1 mA) appears more difficult than a pulsed linac implementation (~ 50mA), again due to the longer injection and storage times associated with the lower current CW beams. The injection foil and beam loss issues are of particular concern for the CW cases, but the resonant foil by-pass scheme may mitigate this issue.

With respect to the choice of ICD-1 and ICD-2 as a basis for extrapolation to a MC proton driver, the ICD-1 is the preferred option. This case requires the least additional upgrades in an accelerator chain that supports a MC proton driver, and minimizes the physics challenges associated with producing the required extremely short bunch structure. With the ICD-2 option (1 mA CW 2 GeV linac, + RCS) neither the NF nor MC needs can be met. Upgrading the ICD-2 linac to a pulsed higher energy linac (6-8 GeV ) may offer some possibility as a proton driver. To this end variable couplers capable of higher peak powers would need to be installed initially to permit reuse of the linac components in the tunnel. Also, if the resonant foil by-pass scheme proves feasible, an energy upgrade of ICD-2 for a MC proton driver may be possible, but this proposal requires more study before adoption.

### Recommendations

Judging the feasibility of either ICD-1 ort ICD-2 as the initial accelerating component of a proton driver supporting the MC collider mission is made difficult by the lack of a proposed configuration that meets the requirements, although ICD-1 has more potential. A self consistent set of accelerator parameters for the proton driver acceleration chain supporting this mission should be assembled. These include:
- the charge exchange injection process: including details of the injection scheme, foil temperature rise and lifetime estimates, and beam loss from foil scattering, and analysis of laser stripping charge exchange injection alternative.
- collective effects: such as longitudinal stability and space charge tune shift effects in the compression stage need evaluation.
- bunch rotation and extraction scheme details.

Finally hardware R&D requirements need to be identified in areas such as:

- Variable power couplers for the linac, which support low current CW operation as well as high instantaneous power in pulsed mode, if the ICD-2 option is pursued.
- RF hardware needs for providing the final bunching and rotation in the compression stage.
- Magnets, as required for the relatively small ring circumferences needed to keep space charge tune shifts low.
- Kickers required for extracting the short bunch lengths.